# Characterization of the structural dynamics of the Aurora-A kinase's activation loop

Maria Grazia Concilio,[a1*] Alistair J. Fielding,[a2*] Richard Bayliss,[b] and Selena G. Burgess[b]

[a]The Photon Science Institute and School of Chemistry, EPSRC National EPR Facility and Service, University of Manchester, Manchester, M13 9PY, United Kingdom.
[b]Astbury Centre for Structural and Molecular Biology, Faculty of Biological Sciences, University of Leeds, Leeds LS2 9JT, United Kingdom.

## Abstract

The understanding of kinase structure is mostly based on protein crystallography, which is limited by the requirement to trap molecules within a crystal lattice. Therefore, characterisations of the conformational dynamics of the activation loop in solution are important to enhance the understanding of molecular processes related to diseases and to support the discovery of small molecule kinase inhibitors. In this work, we demonstrated that long molecular dynamics (MD) simulations exhaustively sampled all the conformational space of the Aurora-A kinase's activation loop and of methane-thiosulfonate spin label (MTSL) attached to the activation loop. MD was used to determine structural fluctuations, order parameters and rotational correlation times of the motion of the activation loop and of the MTSL. Theoretical data obtained were used as input for the calculation of room temperature 9 GHz continuous wave (CW) EPR of MTSL labelled Aurora-A kinase in solution and comparison between simulated and experimental spectra revealed that the motion of the protein and spin label occurred on comparable timescales. This work is a starting point for deeper experimental and theoretical studies of the rotational and translational diffusion properties of Aurora-A kinase related to its biological activity.

---

[1*] First corresponding author details: email mariagrazia.concilio@postgrad.manchester.ac.uk, Telephone +44(0)7769765464.
[2*] Second corresponding author details: email alistair.fielding@manchester.ac.uk, Telephone +44 (0)161-275-4660, Fax +44 (0)161-275-4598.

# 1. Introduction

Site-directed spin labelling (SDSL) combined with electron paramagnetic resonance (EPR) spectroscopy is a very powerful technique used widely for studying the structural properties and dynamical processes of biological systems.[1, 2] This has provided valuable information on many proteins such as T4 Lysozyme, [3] lipoxygenase L-1, [4] α-synuclein, [5] bacteriorhodopsin,[6] SNARE [7] and NavMs. [8] Studies on singly labelled proteins can reveal a wealth of information on the tumbling and diffusion properties of the target by analysis of the continuous-wave (CW) EPR lineshape. [3, 9] A spin label employed to study structural and dynamic properties of biomolecules by EPR spectroscopy is methane-thiosulfonate spin label (MTSL) (Figure 1) that has been widely characterized in previous work [10, 11] and often the label of choice [12] in different biological systems.

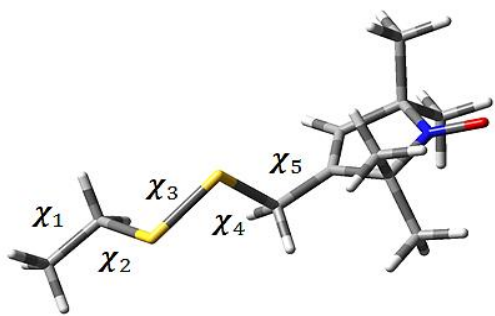

**Figure 1:** Structure of the MTSL side chain with five dihedral angles indicated. The nitrogen and the oxygen atoms of the nitroxide group (NO) are represented in blue and red, respectively.

In order to provide an interpretation of the CW EPR spectrum of MTSL spin labelled systems several different computational approaches to determine the conformational space and dynamics of spin label and subsequently simulate the EPR spectrum have been suggested by several groups. Some approaches are based on Brownian dynamics (BD), introduced by *Robinson et al.* [13] and *Steinhoff et al.* [14] and molecular dynamics (MD) methods introduced by *Budil et al.* [15] and *Oganesyan* [16, 17] to generate stochastic dynamical trajectories of the spin labels and/or to derive diffusion parameters such as the rotational diffusion tensor, diffusion tilt angles and expansion coefficients of the orienting potentials that are then incorporated in the stochastic Liouville equation (SLE) for the calculation of the EPR spectrum. Other methods for the calculation of the EPR spectra include the stochastic Markov models [18-20] and temperature scaling or replica exchange methods, [15, 21] including simulated tempering (ST) and parallel tempering (PT) that have been developed to improve sampling and kinetic information. There are also integrated computational approaches (ICA) that link a quantum mechanical (QM) method rooted on density functional theory (DFT) to the stochastic Liouville equation (SLE) equation in the Fokker-Planck (FP) form. A similar approach was used to calculate the CW EPR spectra of free radicals in their environments [22-24] and to study the structure and dynamics of the MTSL side chain in protein systems, such as T4 lysozyme.[25, 26] The models produced were in good agreement with experimental data with a reasonable computational cost [27-29] and consistent with data obtained by MD simulations. In these approaches QM methods were used to determine EPR parameters (*g*- and *A*-tensors), while coarse grained methods were used to produce a hydrodynamic model of the diffusion tensor [30] that are the input of the SLE equation for the calculation of the CW EPR spectra. [31] There are also computationally inexpensive approaches that can be used to predict the distribution of the conformations of the spin labels and distances between spin-labelled sites, such as MMM, [32] PRONOX [33] and MtsslWizard. [34] There are also MATLAB based EPR simulation software, such as *EasySpin* [35] *and Spinach* [36] that generate spectra in good agreement with those produced by software developed by J. Freed's group implementing the microscopic-order-macroscopic-disorder (MOMD) [37] and slowly relaxing local structure (SRLS) [38] approaches based on the stochastic Liouville equation (SLE) developed by Kubo [39] in the 1963 and then, adapted for EPR simulations by Freed and co-workers for the calculation of slow motion and rigid limit spectra. [40-44]

In this work, we employed molecular dynamics (MD) simulations to characterise the conformational dynamics of Aurora-A kinase's activation loop in solution, a clustering analysis [45] and a principal component analysis (PCA) [46] were performed to determine structure populations and dynamics of the system from MD. Structural fluctuations were characterized through the root mean square fluctuations (RMSFs) and order parameters ($S^2$) calculated using the isotropic reorientational eigenmode (iRED) approach [47] that provided rotational correlation times, $\tau_R$, related to the overall tumbling and reorientational motions of protein. This approach can be applied only to N-H bonds within the protein backbone and is the only method to calculate S values within the AMBER software and was used in this work to determine the mobility of residue 288 within the activation loop of the kinase, where the MTSL spin label was attached to perform EPR studies in solution. The $\tau_R$ values of MTSL were determined through the fit of the auto-correlation function related to the motion of vectors representing its bonds, as performed in previous work. [17, 48] MD were used to determine input parameters (S and $\tau_R$) for the simulation of the room temperature continuous-wave (CW) EPR spectra of spin labelled Aurora-A kinase that showed a lineshape very common in EPR and described in previous work [16, 37, 49, 50] as a superposition of two components; arising from slow and fast motion of the spin label, respectively, and can be simulated using any kind of software package. In this work we reported only 9 GHz EPR measurements since as experiments [3, 19, 51] intended to describe fast motions for this particular protein at 94 GHz EPR spectra produced spectra that were almost insensitive to spin label motion (Figure 1 ESI). A powder pattern was observed even at room temperature at 9 GHz justifying the uselessness of the higher frequencies for the study of the Aurora-A kinase using CW EPR.

Aurora-A kinase is a serine/threonine protein kinase that regulates mitotic entry, centrosome maturation and bipolar spindle assembly and is overexpressed in a number of cancers including breast, colorectal, ovarian, and glioma.[52, 53] Kinase activity is tightly regulated by conformational changes in a conserved region of the protein known as the activation loop upon phosphorylation of threonine residues at positions 287 and 288 and binding of the activator protein Targeting Protein for Xklp2 (Figure 2).[54, 55]

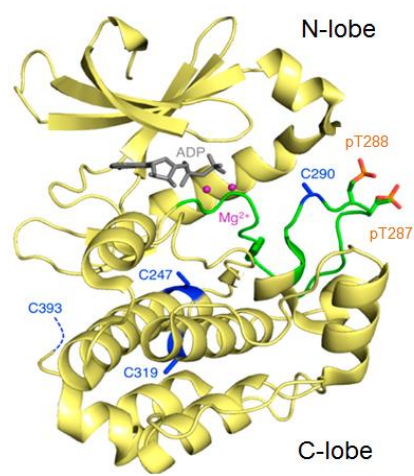

**Figure 2:** Cartoon representation of the kinase domain of Aurora-A (PDB 1OL7 [54]). The activation loop is coloured green (residues 274-299) with the phosphorylated threonine residues at position 287 and 288 shown as sticks. Cysteine residues are shown as blue sticks apart from Cys393, which is missing from the crystal structure. ADP is coloured grey with co-ordinated magnesium ions shown as magenta spheres.

The conformation of the activation loop may also be influenced by the binding of inhibitors to the active site of Aurora-A and, in the case of the potent and selective inhibitor MLN8054, the position of the activation loop main chain is moved by up to 19 Å. [56] The understanding of kinase structure is mostly based on protein crystallography, which is limited by the requirement to trap the molecules within a crystal lattice.[57, 58] Therefore, studies of kinase activation, through characterisation of activation loop conformations in solution are important to enhance our understanding of molecular processes related to diseases and to support the discovery of small molecule kinase inhibitors. In this work, MTSL was introduced into the activation loop of Aurora-A kinase at residue 288. To allow this modification, the wild-type threonine residue was mutated to cysteine, while the threonine at position 287 was mutated to alanine in order to observe conformational changes in a more dynamic activation loop and the solvent exposed cysteine residues at position 290 and 393 were mutated to alanine to prevent labelling with MTSL.

The purpose of this work is not to propose new methodologies, but to combine existing theoretical and experimental approaches in order to characterize the conformational dynamics of the activation loop of Aurora-A kinase, which has never been studied in solution before using EPR spectroscopy. Exhaustively sampling the conformational space of the activation loop and MTSL using free MD simulations is a challenging task due to their weakly structured nature related to inherent flexibility. We also identified several limits of theoretical approaches used in this work that can be suggestion for future work on this topic.

## 2. Methods

### 2.1 Force field parameterization of the MTSL side chain and MD simulation details

The latest AMBER force field [59] (ff14SB), recommended for the study of protein dynamics was extended in order to perform MD simulations with AMBER 2014. [60] The structure of the MTSL side chain was first optimized with the B3LYP hybrid functional [61, 62] using the 6-31G(d) basis set [63]. This level of theory was used as it provided accurate experimental geometries of the MTSL spin label [18] in previous work. Subsequently, in order to reproduce the electrostatic potential and hydrogen-bonding properties of the MTSL side chain, the atom-centred point charges were calculated with Hartree-Fock (HF) theory [64, 65] with the 6-31G(d) basis set, these DFT calculations were performed using Gaussian09 revision d.01.[66] The electrostatic potential was calculated using the restrained electrostatic potential (RESP) procedure. [67] The atom charges determined to extend the AMBER force field to the MTSL side chain were comparable with those reported in literature are shown in Table 1 ESI.[18, 29, 68] To parametrize the NO group, the bond angles, angle bending, regular torsions and non-bond interactions parameters were taken from reference [29].

MD simulations were carried out on MTSL spin labelled Aurora-A kinase, using the X-ray crystal structure of the Aurora-A kinase domain (residues 122-403 C290A C393A; PDB 4CEG [56] with a resolution of 2.10 Å and R-value of 0.202). The pdb file was edited in the website WHATIF [69] to mutate the native Threonine-288 to cysteine. Alanine substitutions were made at residues T287, C290 and C393 in order to produce a structural model similar to the experimentally studied protein. The missing crystallographic hydrogens and MTSL were added using the LEaP module AMBER. The protein was solvated using the Extended Simple Point Charge (SPC/E) water model (11896 water molecules) in a truncated octahedral box with a buffer of 12 Å between the protein atoms and the edge of the box. This water model was used as it predicts viscosities and diffusion constants closer to the experimentally observed data[54] and reproduces crystallographic water positions more accurately than the commonly used Transferable Intermolecular Potential 3 Point (TIP3P) water model. [2, 70, 71] The initial configuration of the MTSL was determined fixing the $\chi_1$, $\chi_2$ and $\chi_3$ equal to $-60^0$, $+80^0$ and $+90^0$, respectively. These values were determined at the minima of the torsional profiles calculated in a previous work [72].

The MD simulation was performed in three main steps: minimization, equilibration and production. Firstly, a short energy minimization was performed in two steps in order to clean the structure and to remove bad contacts using the Simulated Annealing with NMR-derived Energy Restraints (SANDER) module of AMBER. Three chloride ions were used to neutralize the net charge. In the first stage, the water molecules and counter ions were relaxed with 200 cycles of minimization. In the second step, the entire system as a whole was relaxed with 1000 cycles of minimization. Subsequently, the system was heated at constant volume for 20 ps from 10 K to 300 K with 10 kcal/mol weak restraints on the protein. This process was followed by two equilibration steps, the first was performed at constant pressure (1 atm) and temperature (300 K) for 200 ps with no restraints and the second was performed in a microcanonical (NVE) ensemble for 1 ns.

The production step was performed in the NVE ensemble in five steps up to 470 ns collecting time windows equal to 270 ns, 320 ns, 370 ns, 420 ns and 470 ns in order to obtain a good statistical sampling quality. The NVE ensemble was used since the interest of this work is on the dynamics of a system that would be perturbed by the use of a thermostat. This system is large enough (with 14000 atoms in total) that microcanonical and canonical ensemble are almost the same. All the calculations were performed using 1 fs integration time step and coordinates were collected every 2 ps, a total of 235000 frames were collected. The dynamical long-range electrostatic interactions were treated using a particle mesh Ewald (PME) algorithm with default parameters and a 10 Å cut-off Lennard-Jones. A clustering analysis was performed to identify different conformational states sampled during the MD simulation by grouping molecular structures into subsets based on their conformational similarity. The clustering analysis was performed using a version of CPPTRAJ in the AMBER software clustering 5 PC projections by RMSD using the average-linkage algorithm with a sieve of 250 frames. All the MD simulations were performed using one NVidia GTX 980 GPU and with AMBER software optimized to run entirely on a CUDA enabled NVIDIA GPU using a mixed-precision SPDP (single precision, double precision) model that is comparable with the double precision model on a central processing unit (CPU).[73]

**2.2 Calculation of EPR parameters (*g*- and *A*-tensors) and EPR spectra**

The EPR parameters were calculated using the Gauge-Independent Atomic Orbital (GIAO) [74] method, the B3LYP hybrid functional and the latest N07D basis set that has been used with success for accurate calculation of the magnetic tensors ($\Delta g_{ii}$ = ±0.0005, $\Delta A_{ii}$ = ±1 G) in gas phase and in solution of nitroxide radicals at a reasonable computational cost [27, 28] and can be downloaded from the DREAMSLAB website [75]. The Polarizable Continuum Model (PCM) was used to describe solvation in water since the experimental EPR spectrum was measured in water [76-78].

All the EPR spectra were simulated using the open source *Spinach* software library, version 1.7.2996.[36] The room temperature 9 GHz CW EPR spectrum was simulated using the *gridfree* context that uses Fokker-Planck formalism [79, 80] that generates spatial dynamics using lab space rotation generators for the three-dimensional rotational diffusion, $\frac{1}{6\tau_R}(\hat{L}_x + \hat{L}_y + \hat{L}_z)$, where $\tau_R$ is the rotational correlation time and $\hat{L}_i$ represents the angular momentum (not spin) operators acting in the lab space to generate spatial rotations of the spin system from the starting Cartesian coordinates obtained from selected MD frames. The 9 GHz and 94 GHz EPR spectra, measured at 150 K, were simulated using Lebedev spherical powder grids with rank 131.

**2.3 Experimental section**

Spin-labelled Aurora-A 122-403 T287A, T288C, C290A, C393A was produced as stated in earlier work [81]. 50 μM MTSL-Aurora-A 122-403 T287A, T288C, C290A, C393A was used for CW EPR studies. 9 GHz measurements were performed using a Bruker Micro EMX spectrometer at 298 K, the modulation frequency was set at 100 KHz and the microwave power at 2.0 mW. The spin labelling efficiency was equal to 86% as measured following a published procedure. [82] The 94 GHz CW measurements were performed on a Bruker E560 spectrometer. The magnetic field was calibrated using a $Mn^{2+}$ power standard (0.02% MgO) and the procedure described by O. Burghaus *et al*. [83] Dual-scan measurements were made in order to avoid hysteresis effects and a modulation frequency of 100 KHz and low microwave power (0.0048 mW) were used to avoid distortion of the lineshape.

**3. Results and Discussion**

**3.1 Characterization of the dynamics of the MTSL spin label**

During the MD simulation the MTSL fully probed the space around the point of attachment and explored all the regions surrounding the loop. Figure 3 shows the plot of the transitions of the five dihedral angles over the time and a comparison between dihedral transitions and oscillations in the root mean square deviation (RMSD) of the activation loop and the protein.

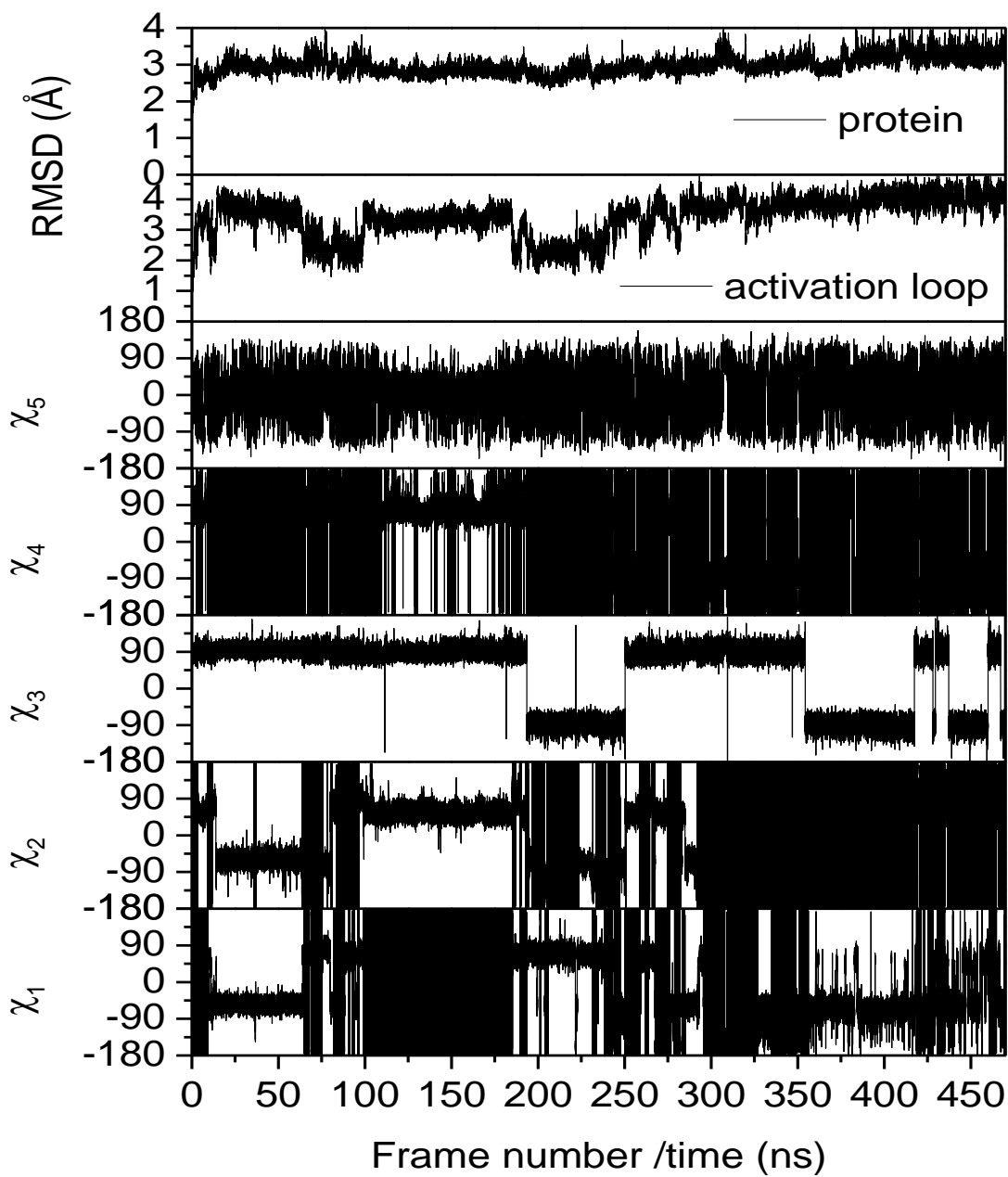

**Figure 3:** Plots of the transitions of the $\chi_1, \chi_2, \chi_3, \chi_4$ and $\chi_5$ dihedral angles and of RMSD of the protein and of the activation loop and over 470 ns. The RMSDs were calculated considering the Cα, C, N, O atoms and the reference frame was the first frame after the equilibration step.

It was observed that the RMSDs of the activation loop and the protein reasonably converged at ~3-4 Å, indicating not a large variation from the initial X-ray crystal structure. The RMSD of the protein remained constant ~3 Å, while the RMSD of the activation loop was characterized by several fluctuations that corresponded to simultaneous transitions of $\chi_1$ and $\chi_2$ between 10 ns and 64 ns, and between 98 ns and 185 ns. After 250 ns, the RMSD of the activation loop became stable and less transitions of $\chi_1$ and $\chi_2$ were observed, this indicated a coupling between transitions of the dihedral angles and oscillations of the protein backbone, no evidences of correlations between transitions of $\chi_4$ and $\chi_5$ and the oscillations of the RMSDs were found. This indicated no correlation between the backbone dynamics and the motion of the N-O group in the pyrroline ring. The coupling between transitions of dihedral angles close to each was also observed, *e.g.* the simultaneous transitions of $\chi_1$, $\chi_2$ and $\chi_3$ dihedral angles between 192 ns and 249 ns, and $\chi_4$ and $\chi_5$ between 108 ns and 170 ns.

The plot of the transitions of the five dihedral angles over the time showed slow transitions of $\chi_1$ and $\chi_2$ due to high energy barriers of 4-6 kcal mol$^{-1}$ separating the conformational minima in the torsional energy profiles observed in previous work [72]. Transitions of $\chi_3$ were very slow and less frequent due to higher energy barriers of 20 kcal mol$^{-1}$, while transitions of $\chi_4$ and $\chi_5$ were fast due to low energy barrier of 1-2 kcal mol$^{-1}$. A comparison between the torsional energy profiles of the five dihedral angles calculated in previous work [72] and their transitions (Figure 3) revealed that the values of the minima determined by QM calculations were well reproduced by MD simulations. Three minima were found in the torsional profile of $\chi_1$ corresponding to -160°,-60° and +60° and the same three minima were found in the plots of the Gibbs free energy surfaces determined from MD (Figure 2 ESI) from which the minima of $\chi_2$ were confirmed and so on for $\chi_3$, $\chi_4$ and $\chi_5$. A comparison between the Gibbs free energies determined MD and DFT (Figure 2 and Table 2 in ESI) considering ten conformers of the MTSL revealed a difference of ~5.4 -5.8 kcal mol$^{-1}$ and conformers characterized by comparable free energy. The plots of the Gibbs free energies obtained from MD revealed a small difference of ~4.2 kcal mol$^{-1}$ between high and low energy regions indicating that the rotamer population is expected to be fully sampled at room temperature and each conformation has the same overall orientational probability. In order to determine the dynamics of the MTSL, the auto-correlation functions of vectors connecting two atoms of the spin label (the full list is shown in Figure 2 and Table 2 in ESI) were calculated within the AMBER software using spherical harmonics with rank equal to 2, as described in [17, 84], corresponding to $\langle D_{00}^2(\theta(0))D_{00}^2(\theta(t))\rangle$ with $D_{00}^2 = (3\cos^2\theta(t) - 1)/2$ and $\theta(t)$ representing the variation in the time *t* of the direction of the vector. The auto-correlation functions showed a multi-exponential behaviour, indicating different time scales of the motional contributions and were fitted using a bi-exponential function from which two correlation times, $\tau_1$ and $\tau_2$ were extracted. The fit to the auto-correlation function of the vector C$\alpha$-C3 was not feasible due to the trend of its curve and unreasonable results were obtained for the vector C3–S1. Values of $\tau_1$ obtained for the vectors connecting atoms of the MTSL side chain were seen to be very different from each other and oscillating between 10 ns and 57 ns, while quite comparable values of $\tau_2$ were obtained that decreased from 3.7 ns for the vector C3–S1 (close to the point of attachment of the spin label) to 2.7 ns for the vector C4-C5 (close to the pyrroline ring), in good agreement with the plots of the transitions of the dihedral angles that showed very slow motion about $\chi_3$ and fast motions about $\chi_4$ and $\chi_5$. The fits to the auto-correlation functions of the vector related to the N-O group revealed values of $\tau_1$ and $\tau_2$ smaller than those observed for the other vectors, indicating that the motion of the N-O in the pyrroline ring is uncoupled from the motion of the spin chain. The two different correlation times obtained from the auto-correlation function can represent slow and fast rotameric transitions of the MTSL spin label. MD revealed that both configurations of the MTSL were fully exposed to the solvent (Figure 4A) and interacting with residues within the N- and C-lobes of the protein (Figure 4B and C).

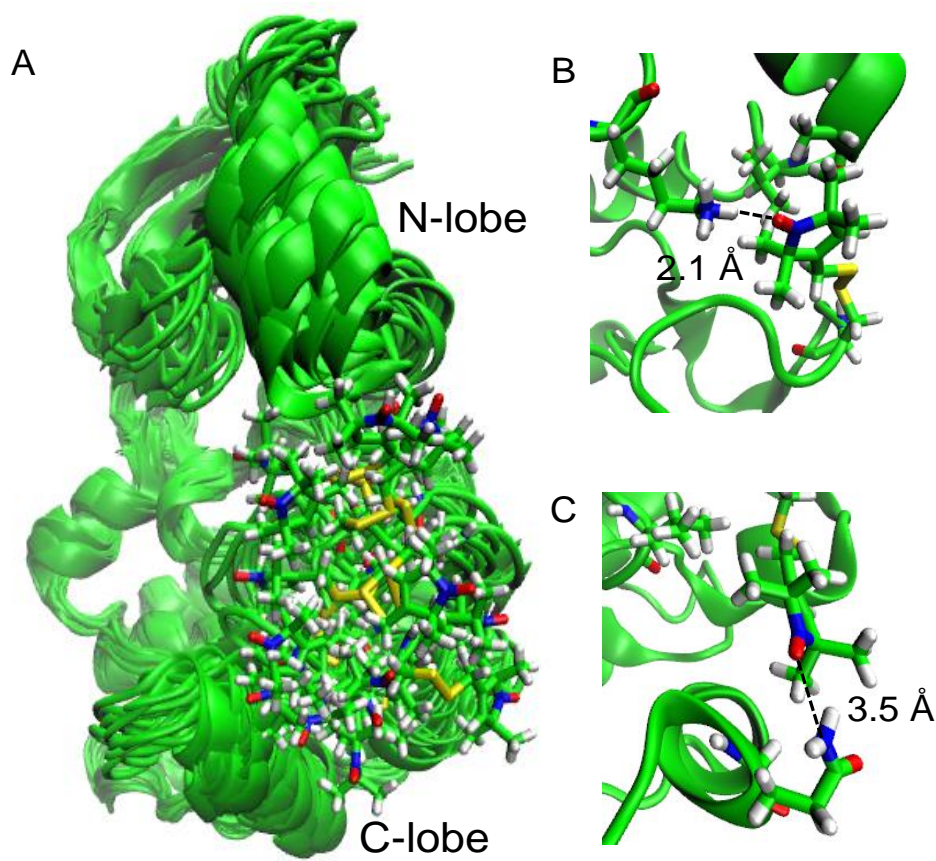

**Figure 4:** (A) Distribution in the space of the MTSL spin label obtained from the MD simulation, (B) Configuration of MTSL interacting with the lysine 46 N-lobe of the protein. (C) Configuration of MTSL interacting with the glutamine 209 in the C-lobe of the protein.

An isotropic and uniform distribution of conformations that remained constant in space and time was observed, similar results were obtained from the MMM rotamer library (Figure 4 ESI). Faster transitions can be associated to configurations of the spin label fully exposed to the solvent, while slower transitions can be explained considering interactions between atoms of the spin label and the residues of the protein, however, a large exposure to the solvent was observed in all the possible configurations giving evidence of no restricted motions.

### 3.2 Characterization of the conformational states and dynamics of Aurora-A's activation loop

The clustering and PC analyses were performed to identify the conformational states and to determine the five most dominant modes of motions of the activation loop. The clustering analysis was performed on five different regions of the full MD trajectory, specifically between 0 ns and 270 ns, between 270 ns and

320 ns, between 320 ns and 370 and between 420 and 470 ns. Figure 5 A shows the representative structures obtained with lowest RMSD to the centroid in the ten most populated clusters determined in the five regions.

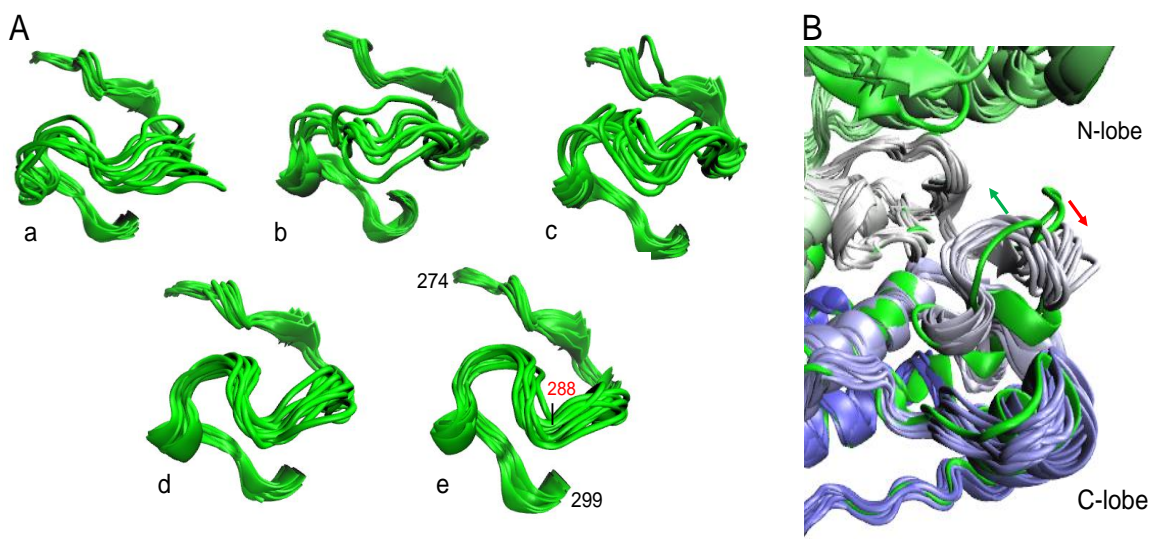

**Figure 5:** Comparison between conformational states of the Aurora-A kinase activation loop (from residues 274 to 299) obtained from the MD simulation. (A) Representative structures obtained from the clustering analysis at different time windows, (a) 270 ns, (b) 320 ns, (c) 370 ns, (d) 420 ns and (e) 470 ns. The position of residue 288 is indicated in red. (B) Comparison between the energy minimized X-ray structure of Aurora-A (green structure) and the 10 representative structures obtained after 470 ns (white structures). The green and the red arrows indicate the differences between the initial X-ray crystal structure and those obtained in the final region of the MD simulation.

The activation loop was seen to adopt different conformational states in the time windows at 270 ns, 320 ns and 370 ns, while no significant structural changes were observed in the time windows at 420 ns and 470 ns, indicating less dynamics on longer timescales. In the time window between 420 ns and 470 ns, high values of the pseudo F-Statistic (pFS) and the Davies-Bouldin Index (DBI) were obtained and equal to 1731 and 1.55, respectively, indicating a good quality of cluster analysis. Figure 5B shows a comparison between the ten calculated conformations of the activation loop observed in the last MD time window (420 ns - 470 ns) and the energy minimized X-ray structure. The average RMSD was equal to 3.1 Å, indicating no significant changes from the starting structure. However, it was observed that residues from 289 to 295 moved inward toward the N-lobe of the protein, while residues from 288 to 282 moved toward regions more exposed to the solvent. Figure 6 shows five molecular structures obtained from the PC analysis performed in the last MD time window and the predominant modes of motion (indicated by the arrows) observed within the activation loop and the plot of the root mean squared fluctuations (RMSF) for each active mode.

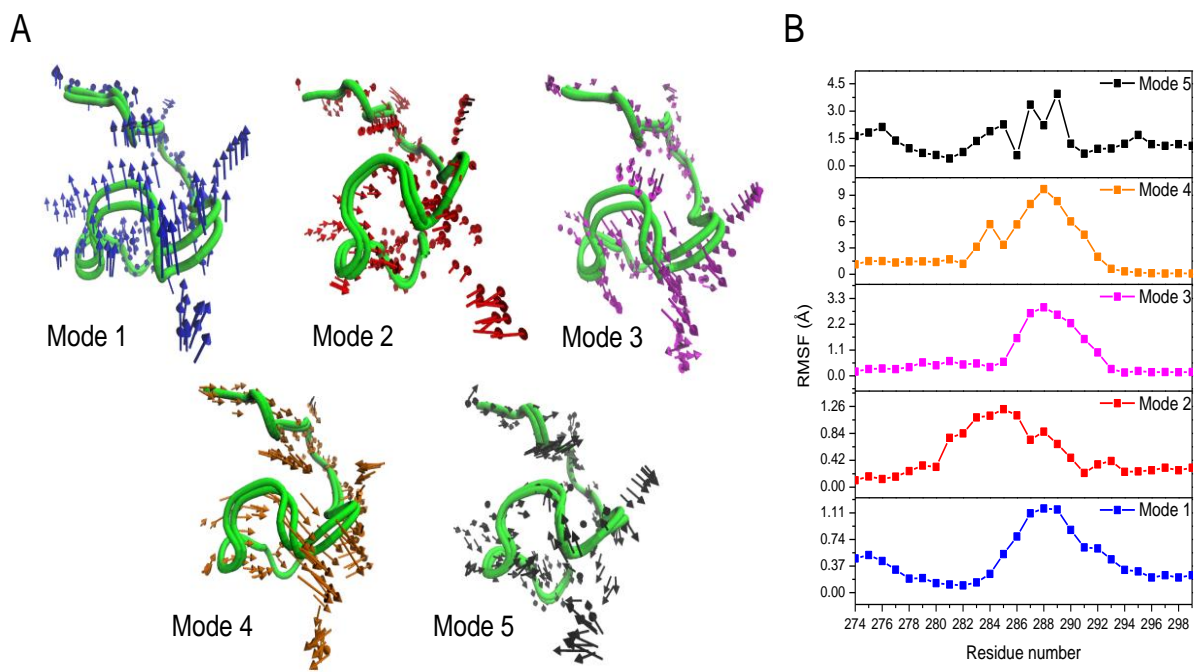

**Figure 6:** Molecular structures of the activation loop of Aurora-A kinase obtained from the projected frames of the first, second, third, fourth and fifth PC mode. The arrows represent the direction of the motions of the activation loop and the length represents the magnitude of the oscillations. (B) Corresponding plots of the RMSFs for each active mode.

The five PCs corresponded to global protein bending motions (amination of these motions, Modes 1-5, are available in Supporting Information) going from the smaller fluctuations ~1.1 Å observed in Mode 1 to the wider fluctuations ~ 4.5 Å and 9.0 Å observed in Mode 4 and 5, respectively. The plots of the RMFS (Figure 6B) showed large fluctuations, indicated by sharp peaks, in the region between residues 284 and 294 within the activation loop, while much smaller fluctuations were observed for the other residues of the activation loop. Sharper peaks were observed at identical positions indicating concentred fluctuations within the modes. In order to measure how well the motions converged to each other, the histograms of the PC projections were determined and significant overlap was obtained (Figure 5 ESI). Higher fluctuations were observed for modes 3, 4 and 5 indicating higher flexibility when compared with modes 1 and 2, mode 4 was the most flexible. In Figure 6A, it is possible to observe that the residues of the activation loop fluctuated in the same direction for modes 1, 2, 3 and 4, while fluctuations in different directions were observed for mode 5.

In order to determine timescales of the motions within the activation loop, the isotropic iRED analysis [47] was performed along the full MD trajectory in order to determine eigenmodes and amplitudes corresponding to tumbling and reorientational motions of activation loop. This approach incorporates motional correlations between reorientations of backbone N-H bonds in the form of a covariance matrix analysis of internuclear vector orientations that were represented by spherical harmonics of rank 2. In the iRED analysis, the correlation times $\tau_m$ of protein motions, along the reorientational eigenmodes, can be estimated from the time-correlation functions $C_{m,l} = \langle a_{m,l} * (\tau + t) a_{m,l}(\tau) \rangle_\tau$ calculated for each mode, where $a_{m,l}$ are the time-dependent amplitudes obtained by projecting MD snapshots on the eigenmodes calculated from corresponding to IRED vectors that were set equal to 26, corresponding to the number of residues of the activation loop (the number of defined IRED vectors have to match the number of eigenmodes calculated). The corresponding $\tau_m$ values of these correlation functions can be determined by exponential fitting using the equation $\tau_m \cong \frac{1}{C_m(0) - C_m(t \to T)} \int_0^T C_m(0) - C_m(t \to T) \, dt$, where $C_m(t \to T)$ indicates the plateau value of $C_m(t)$. The iRED approach was used also to calculate the order parameters $S^2$ of each residue of the activation loop in order to determine the local rigidity. These values corresponded to the amplitude of motion within a cone formed by bond vectors (N-H bonds within the activation loop) such that with increasing rigidity, a smaller cone angle $\Omega$ would be produced. Site-specific values of $S^2$ can range between 0 for a completely disordered bond-vector, to 1 for a completely rigid bond-vector. [85] Figure 7 shows the $C_m(t)$, the corresponding $\tau_m$ calculated for each of the 26 modes and the order parameter of each residue of the activation loop calculated with $S^2 = \langle P_2(cos(\Omega_i - \Omega_j)) \rangle_{ij}$, where $P_2$ is the Legendre polynomial of rank 2 and $\Omega_i - \Omega_j$ is the orientational change of the internuclear vector when going from snapshot $i$ to snapshot $j$.

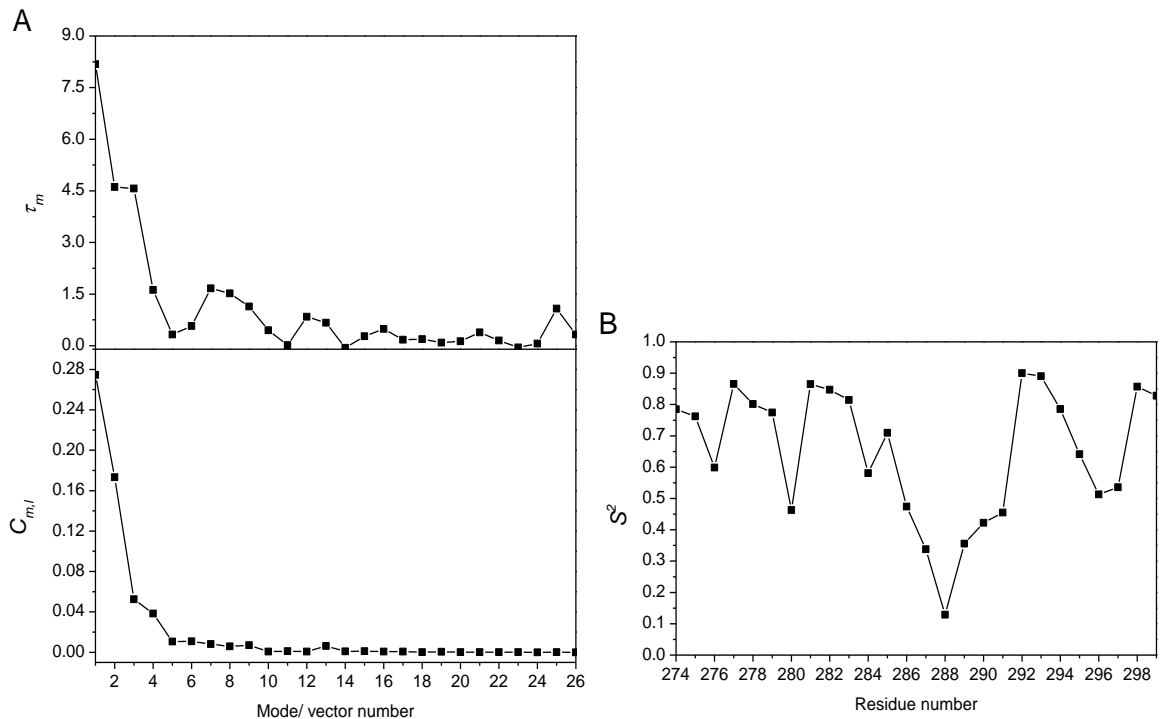

**Figure 7:** Results obtained from the iRED analysis performed along the full MD trajectory. (A) Time-correlation function, $C_{m,l}$ and corresponding correlation times, $\tau_m$ for each mode/iRED vector. (B) Order parameters, $S^2$ for each residue of the activation loop.

The modes decay in good approximation, monoexponentially, and corresponded to individual reorientational modes of motion and rotational tumbling. The extracted correlation times varied between 0 ns and 8 ns, indicating a wide distribution of motions within the activation loop and structural fluctuations dominated by slower time-scale events. The calculated $S^2$ for the residues of the activation loop varied between 0.75 and 0.13, the less flexible regions were observed between residues 274 - 283 and 292 - 299 where $S^2$ varied between ~0.85 and ~0.50. This is consistent with the plots of the RMSFs (Figure 6) that revealed higher flexibility between residues 284 and 292, where the most flexible residue is 288 ($S^2$ = 0.13) that corresponds to the point of attachment of the MTSL spin label and due to its length and flexibility increased the mobility of the residue.

### 3.3 Description of the 9 GHz CW EPR spectra of Aurora-A kinase

The room temperature CW EPR spectrum of Aurora-A kinase showed a lineshape very common in EPR and described in previous work as a superposition of two components obtained using two different rotational correlation times, one arising from configurations of the spin label in restricted states and the other arising from configuration of the spin label highly exposed to the solvent. [16, 37, 49, 50] In this work, the principal values of *g*- and *A*-tensors were determined from the simulation of the 9 GHz EPR spectrum (Figure 8A), measured at 150 K to freeze out all motion and were equal to $g_{xx}$=2.0086, $g_{yy}$=2.0065, $g_{zz}$=2.0029; $A_{xx}$= 4 G, $A_{yy}$= 6 G and $A_{zz}$= 36 G. These values are in good agreement (*Δg* ~ ±0.0005 and *ΔA* ~ ±2 G) with those obtained from 94 GHz measurements (Figure 6 ESI) and DFT calculations performed using an energy optimised structure of the MTSL like shown in Figure 1. The 9 GHz EPR spectrum of Aurora-A kinase (Figure 8) was fitted using two components in a 1:1 ratio two spectra obtained with $\tau_R$= 3 ns and $\tau_R$= 11 ns, indicating that 50 % of the spin label exhibited slow dynamics and the remaining 50% exhibited fast dynamics. A low amount of free spin label equal to 14 % was added to the spectrum in Figure 7B to reproduce the sharp component at 335 mT.

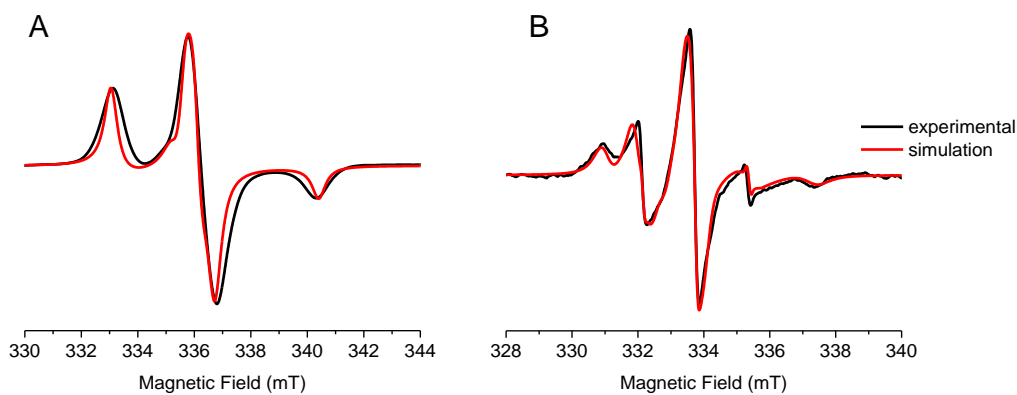

**Figure 8:** Comparison between experimental and simulated 9 GHz EPR spectra of spin labelled Aurora-A kinase measured at 150 K (A) and at 298 K (B), respectively.

In this work the best fits were obtained using $S^2$ equal to zero that is very close to the $S^2$ = 0.13 determined by the iRED analysis (Figure 7B), indicating a low restriction of the N-H bond and high flexibility of the residue 288. This was also observed in the plot of the RMSFs in Figure 6B. However, remembering the definition of order parameter that corresponds to the amplitude of motion within a cone formed by a bond vector, an appropriate description of the restriction of the motion of the spin system would require the calculation of the order parameter for the N-O bond, the iRED approach within the AMBER software can be applied only to N-H bonds and therefore only qualitative descriptions of the EPR spectra can be provided. Figure 6 shows that the spin label is largely exposed to the solvent with no evidence of restriction, the fast motion observed in the EPR spectrum can arise from configurations of the spin label freely rotating in the solvent (Figure 6A), while the slow motions can arise from configurations in which the motion is slowed down by interactions with the protein (Figure 6B and C). Values of $\tau_1$ equal to 15 ns and $\tau_2$ equal to 1 ns determined from the auto-correlation function of the N-O vector were very close to those required to simulate the EPR spectrum, indicating that the EPR spectrum can arise from the single motion of the NO group. However, also values of $\tau_2$ between 2 ns and 3 ns, determined for the vectors of the side chain, were in good agreement with that those required to simulate the fast component in the EPR spectrum, while values of $\tau_1$ were much higher than those required. This indicated that the interpretation of the exponential decay of the auto-correlation function might not be valid to describe rotational diffusion in the case of restricted motion and other approaches have to be developed. However, a firm conclusion cannot be given due to the lack of literature of a similar analysis performed on all the vectors of the side chain and further investigations are required in future works.

Values of the correlation times of the reorientational motion of the activation loop, determined from the iRed analysis (Figure 7), were equal to 8.1 ns, 4.5 ns, 1.5 ns and 0 ns, considering an average of 3.5 ns that is very close to correlation time of the motion of the MTSL spin label, we can conclude that the motion of the protein and spin label occur on the same timescale. The 11 ns requested to simulate the slow component of the EPR spectrum can arise from a motion of the spin label slowed down by interactions with residues of the protein. However, a quantitative description of the effect of the protein environment on the spin label would require a description of the potential energy surface of the motions of the spin label in the protein environment that is a complicated function of space and time. Order parameters have to be determined both at the point of attachment of the tether and on its top in order to determine the degree of restriction of the N-O bond by the protein environment. This would require the development of a new approach that would extend the iRed routine to MTSL side chain in order to provide input parameters for the simulations of the EPR spectra; this can be a suggestion for future work.

## 4. Conclusions

In this work we characterized the structural dynamics of the activation loop of the Aurora-A kinase, the clustering analysis revealed different conformational states between 0 ns and 370 ns, after 370 ns conformational states converged to comparable structures indicating less dynamics on longer time scales. A comparison between conformations obtained from MD and the initial X-ray crystal structure revealed small differences, residues from 289 to 295 moved inward toward the N-lobe of the protein, while residues from 288 to 282 moved toward regions more exposed to the solvent, providing new insight useful for the development of new kinase inhibitors. The PC analysis revealed the amplitude of the fluctuations of the residues of the activation loop, the most flexible region included residues between 284 and 292 where RMSFs oscillated between 1.1 Å and 9.0 Å. This result was confirmed by the iRED analysis that provided lower values of the ordering parameters in this region as compared with the other regions of the activation loop.

We also demonstrated that long MD simulations can provide exhaustive sampling of the MTSL rotamers, since multiple transitions of all the dihedral angles were observed, including those of the $\chi_3$ dihedral angle that are not straightforward to obtain. The analysis of the auto-correlation function was performed using different vectors within the MTSL spin label and revealed the presence of slower and faster motions corresponding to configurations of the spin label fully exposed to the solvent and other interacting with other residues of the protein. This result was consistent with the room temperature 9 GHz CW EPR spectrum that was simulated summing two components with two different correlation times equal to 3 ns and 11 ns in good agreement with those determined from the auto-correlation function of the N-O bond, indicating the EPR spectrum strongly depends on the motion of the N-O group in the MTSL side chain. Values of $\tau_2$ extracted from the auto-correlation function related to vectors of the side chain oscillated between 2 ns and 3 ns, while the values of $\tau_1$ oscillating between 20 ns and 57 ns were seen to be too high to simulate the EPR spectra. This indicated that the auto-correlation function might be not valid to describe rotational diffusion in the case of restricted motion and further investigations are required. The iRED revealed correlation times that ranged between 8 ns and 0 ns, indicating a wide distribution of motions within the activation loop and structural fluctuations dominated by events on a slower time-scale that was comparable to that related to the motion of the MTSL. In this work we provided an overview of the structure and dynamics of the Aurora-A kinase activation loop and a next step for the testing of inhibitors and binding partners on the 9 GHz CW EPR spectrum of Aurora-A kinase.

**Author contribution**

**Acknowledgments**

This work was supported by a studentship from Bruker Ltd. to M.G.C. and by a Cancer Research UK grant (C24461/A12772) to R.B. The authors would like to acknowledge the use of the EPSRC UK National Service for Computational Chemistry Software (NSCCS) and its staff (Dr. Alexandra Simperler and Dr. Helen Tsui for some technical advice) at Imperial College London in carrying out this work and the EPSRC National EPR Facility at the University of Manchester. This work is a result of many discussions within the EPR and computational communities. M. G. C. acknowledges Dr. A. Baldansuren and Prof. D. Collison for useful discussions about the EPR lineshape analysis; Dr. N. Burton and Dr. A. Simperler for useful feedback.